\documentclass[11pt]{scrartcl}
\usepackage[final]{allergy}
\usepackage{helvetic}

\usepackage{authblk}

\usepackage{mathrsfs}

\newcommand{\BOSZ}{\textit{BOSZ }}

\author[a]{Artur Matysiak}
\author[b,c]{Volker Roeber}
\author[c]{Henrik Kalisch\footnote{henrik.kalisch@uib.no}}
\author[a]{Reinhard K\"{o}nig}
\author[e]{Patrick J.C. May\footnote{p.may1@lancaster.ac.uk}}

\affil[a]{Research Group Comparative Neuroscience, Leibniz Institute for Neurobiology, Brenneckestr. 6, 39118 Magdeburg, Germany}
\affil[b]{Universit\'{e} de Pau et des Pays de l’Adour, E2S-UPPA, chair HPC-Waves, SIAME, All\'{e}e du parc Montaury, 64600 Anglet, France}
\affil[c]{Department of Oceanography, University of Hawaii at Manoa, 1000 Pope Road, Honolulu, HI 96822, USA}
\affil[d]{Department of Mathematics, University of Bergen, Postbox 7800, 5020 Bergen, Norway}
\affil[e]{Department of Psychology, Lancaster University, Lancaster LA1 4YF, United Kingdom}

\date{}

\title{Listen to the Waves: Using a Neuronal Model of the Human Auditory System to Predict Ocean Waves}

\begin{document}

\maketitle

\vskip -0.3in

\begin{abstract}
\noindent
{\textbf{Abstract.}}
Artificial neural networks (ANNs) have evolved from the 1940s primitive models of brain function to become tools 
for artificial intelligence \cite{1}. They comprise many units, artificial neurons, interlinked through weighted 
connections. ANNs are trained to perform tasks through learning rules that modify the connection weights. 
With these rules being in the focus of research, ANNs have become a branch of machine learning developing 
independently from neuroscience. Although likely required for the development of truly intelligent machines, 
the integration of neuroscience into ANNs has remained a neglected proposition \cite{2}. 
Here, we demonstrate that designing an ANN along biological principles results in drastically 
improved task performance. As a challenging real-world problem, we choose real-time ocean-wave prediction
which is essential for various maritime operations. Motivated by the similarity of ocean waves measured 
at a single location to sound waves arriving at the eardrum, we redesign an echo state network 
to resemble the brain's auditory system.  
This yields a powerful predictive tool which is computationally lean, robust with respect to network parameters, 
and works efficiently across a wide range of sea states. Our results demonstrate the advantages of 
integrating neuroscience with machine learning and offer a tool for use in the production of green energy 
from ocean waves. 
\end{abstract}

\section{Artificial neural networks.}
Functionality is bestowed upon an artificial neural network (ANN) through connection weights that optimize 
the input-output mapping of the network in a task-specific way. The primary computational cost of using ANNs 
therefore arises from the need to change the connection weights according to various learning rules. 
Favored by the increasing availability of scalable computing resources, larger networks with many layers, 
so called deep neural networks, are now routinely employed to deal with complex tasks 
and very large data sets \cite{3,4}. 
However, computational cost remains an issue of practical concern. For example, while it may be possible 
for a proof-of-concept study to use a deep neural network running on hundreds of CPUs, the same cannot be done 
for a system deployed, for example, in an autonomous car or in a drone. Computational cost needs to be kept low 
while achieving high accuracy if real-time operational support is to be provided for stabilization or navigation. 
In an effort to improve both accuracy and efficiency, neural networks have recently 
been coupled with various physical laws, encoded through partial differential equations \cite{5,6}. 
For example, encoding Galilean invariance as a multiplicative layer 
in a neural network enabled a realistic model of turbulent flows featuring rotational invariance \cite{7} 
which is fundamental in any turbulence closure, but not naturally respected by a neural network model. 

Here, we demonstrate a different route to ANN efficiency. We return to the original spirit of ANN development 
by adapting results from neuroscience in network design and operation. 
As an example of a challenging real-world problem, we take time series prediction, 
specifically the generation of phase-resolved forecasts of ocean surface elevation in shallow water. 
Real-time prediction of ocean waves is essential in marine operations and for energy production at sea \cite{8}. 
In particular, the viability of wave power installations using designs such as arrays of point absorbers, 
flap-type converters, or oscillating water column converters depends on maximizing the efficacy of the devices 
with respect to real-time wave conditions. Indeed, it is well understood that the energy take-up 
of wave energy converters (WECs) depends critically on the oscillating structure being 
in resonance with the incoming ocean waves. In many cases, this alignment can be achieved 
using active hydraulic or electro-magnetic control of the device. 
The control algorithm depends on information about the incoming wave profile, and there are a number of methods
which can be used to obtain this information. Since regulating the WEC appropriately requires wave-by-wave information, 
phase-resolving wave modeling should be used. However, standard phase-resolving models are computationally 
too costly to be useful in an operational fashion in this situation. As an alternative, wave measurements by radar, 
survey buoy, or pressure gauges could be performed continuously during operation. Radar or Lidar measurements 
over a large swath of the sea may still necessitate propagating the waves to the location of the WEC numerically. 
On the other hand, measurements in close proximity to the device requires prediction of the signal 
for up to $60$ seconds into the future in order to allow for control adjustment. Various strategies exist 
for dealing with this problem \cite{9}. In deep water, it was shown that standard Auto Regression (AR) signal analysis 
methods are optimal, and that ANN methods are not needed \cite{10}.
In shallower water, however, where the waves exhibit strongly nonlinear behavior, 
ANN methods are expected to outperform standard linear signal analysis techniques. 
Hence, in this work, we focus on wave-by-wave predictions in shallow water.

As our starting point, we use an echo state network (ESN), a computationally lean approach 
for predicting dynamical time series \cite{11}. In ESNs, the input is fed into an ANN - the reservoir - 
which excites an output unit. Contributing to a low computational cost, 
the internal weights in the reservoir are kept fixed and only the output weights are updated during training. 
For this purpose, regression is used for minimizing the error between the output signal 
and the training signal. As a result of training, the output of the network traces the training signal. 
The input is then switched off and replaced with the output signal, at which point the ESN starts 
to generate a prediction of the time series. The term “echo” refers to the state of the reservoir 
encoding not just the incoming signal but also the signal’s recent history. In the current study, 
we refashion the ESN to make it resemble the brain’s auditory system, 
thereby strongly improving predictive performance.


\section{Redesigning the echo state network to resemble the auditory system}
While standard ESNs (stdESNs) rely on a reservoir of randomly connected units receiving the input signal directly, 
in the present contribution, we suggest an alternative approach where the reservoir is structured 
like the auditory cortex and the input is preprocessed similarly as in the cochlea of the ear. 
This approach is motivated by two considerations. First, as its input, the auditory system 
deals with streams of one-dimensional time series: the sound waves hitting each ear drum. 
The situation is therefore analogous to dealing with real-time ocean elevation 
data measured at discrete locations: in both cases, a one-dimensional time series 
is sampled at a single location. Second, the sound wave vibrations of the present moment 
only make sense when perceived against the backdrop of what came before in a time window 
stretching seconds into the past. 
A case in point is the perception of speech and music which implies that the brain 
must represent sounds in the context of a continuously updated representation of the recent past. 
This is indeed what we find in auditory cortex, where responses show strong context effects \cite{12,13,14}. 
This kind of mixing of the current signal with the past encoded in the network state is precisely what an ESN does.

In the auditory system \cite{15}, the three-dimensional sound wave field 
is sampled at the eardrum whose vibrations are then amplified by the middle ear 
and transduced into electro-chemical signals in the cochlea of the inner ear (Fig. 1A). 
The cochlea carries out a spectral analysis of the sound signal, transforming its representation 
into a frequency mapping, the tonotopic map. Here, each neuron is maximally responsive 
to a specific frequency, and the best frequencies are systematically organized along 
one spatial dimension of the nerve tissue. Thus, the spatial pattern of cells 
responding to a sound encodes the frequency content of that sound. The tonotopic mapping is retained 
at each stage of the auditory pathway as neural activity makes its way through the brainstem 
and the midbrain to cortex. The auditory cortex comprises several tonotopically 
organized fields which differ in number, size, and connectedness across species. Despite this variability, 
there is a shared general hierarchical organization (Fig. 1B): At the center are the core fields, 
which receive the subcortical input from the auditory pathway, specifically from the medial geniculate nuclei 
of the thalamus. The signal then propagates to a set of belt fields surrounding the core, 
and from there to parabelt fields. 
This forward spread of activity is carried by feedforward connections, 
and these are complemented by feedback connections. 
%
%
\begin{figure}[t]
\centering
\includegraphics[width=0.9\textwidth]{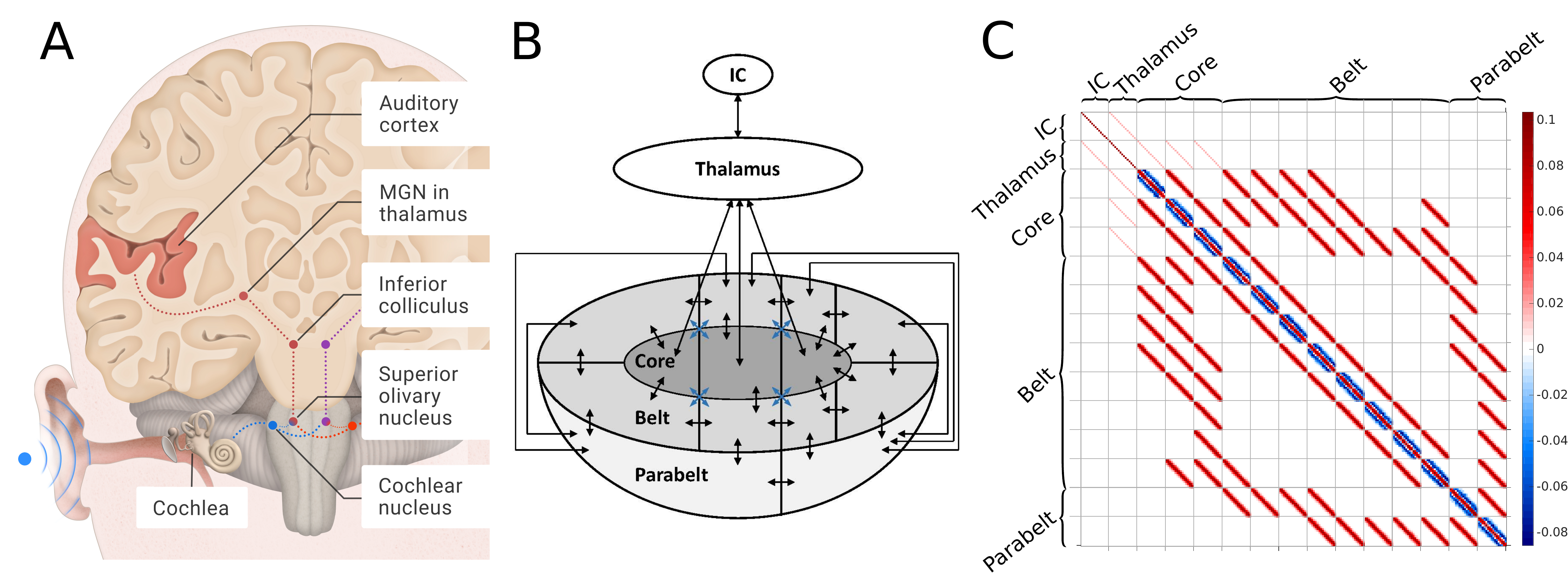}
\caption{\small
Modeling the auditory cortex. (A) In the auditory system, sound waves are transduced 
in the cochlea into electric signals. These propagate up the auditory pathway 
to the auditory cortex via the cochlear nucleus and the superior olivary nucleus 
in the brainstem as well as the inferior colliculus (IC) and the medial geniculate nucleus (MGN) 
in thalamus as part of the midbrain. (B) Schematic view of the primate auditory cortex \cite{16} 
and the two midbrain components. The input from the ascending auditory pathway, i.e. 
from IC and thalamus, arrives in the core area comprising three tonotopically organized fields. 
The core fields are surrounded by eight belt fields and two parabelt fields. 
The black and blue arrows indicate reciprocal connections between fields, 
with high or low density of connections, respectively. 
(C) The structure of the auditory cortex in (B) is represented as a connection matrix, 
which can be used to simulate the functioning of auditory cortex. 
In this model, each field comprises multiple units, and each dot in the matrix representation 
shows a connection between two units. The superposition of two matrices indicating 
excitatory-to-excitatory connections and excitatory-to-inhibitory connections are shown \cite{17} 
(for details, see appendix A). The color bar on the right represents the strength 
of the connections between any two columns: positive values (in red) stand for excitatory, 
negative values (in blue) for inhibitory connections (adapted from Hajizadeh et al. \cite{17}).}
\label{fig1}
\end{figure} 
%
%

The above aspects are captured by a computational model of auditory cortex  \cite{17,18,19,20} 
which we use as the basis for designing the reservoir of the ESN (see appendix A). 
The basic unit in the model is a representation of a cortical column comprising two mean-field neurons \cite{21}, 
one excitatory, the other inhibitory. A key feature of the model is its anatomical structure 
which mimics the hierarchical core-belt-parabelt organization of the auditory cortex. 
This structure is similar among mammals although the number of fields and their connectivity 
is species-specific \cite{22,23}. 
The dynamics of the model are described by pairs of coupled differential equations 
for the state variables of the excitatory and the inhibitory neurons (Eq. \eqref{eq:uv}). 
The column units are connected according to a matrix (Fig. 1C) 
which roughly replicates the connectome of the primate auditory cortex \cite{16}.  
The diagonal stripes indicate tonotopic organization; 
the connections above and below the main diagonal are feedback and feedforward connections, respectively. 
This model accounts for the event-related field \cite{17,20} 
as well as for a plethora of experimental observations 
such as the forward enhancement of neural responses \cite{12}, 
the suppression of neural responses caused by stimulus repetition \cite{13}, 
and enhanced responses elicited by unpredictable stimuli  \cite{14}. 
Observations such as these show that, as a general principle, 
the response to the stimulus represents an interaction between the sensory inputs 
and the internal state of the cortex which encodes the recent past \cite{24}.

In the auditory-cortex-based ESN (audESN), the input signal first passes through a filter bank realized 
by a Fast Fourier Transform (FFT), thus mimicking the spectral transformation of the signal in the cochlea. 
The real part of the FFT is used as input to the excitatory units of the AC network, 
while the imaginary part acts as input to the inhibitory units, thus reflecting the low-frequency and phase-dependent 
nature of ocean waves. The resulting tonotopic representation of the signal is then fed into the reservoir
which is a modification of the computational model detailed above. 
While the full model of the auditory cortex is fairly 
complex \cite{17,18,19,20}, a simplified version is used in the current computations. Here, the units in
audESN are organized into a core field, which receives the pre-processed input and distributes the signal to three belt 
fields surrounding the core field (Fig. 2A). Fields are reciprocally and topographically connected with each other, 
preserving the tonotopic mapping produced by the pre-processing (Fig. 2B). However, connections are partly random 
so that the precise pattern of connections between two fields is unique to that pair. Within each field, 
columns are connected to each other via short-distance inhibitory (Fig. 2B, right panel) 
and longer-distance excitatory lateral connections (Fig. 2B, left panel). 
Figure 3 contrasts the structured reservoir used in audESN (right panel) 
against a randomly connected reservoir used in a stdESN (left panel). 
In the stdESN, the incoming signal expressed by the blue waveform on top of that figure 
is fed directly into all units. In an audESN with connections between columns as introduced in Fig. 2, 
the incoming signal is first pre-processed by a Fast-Fourier-Transform (FFT) 
before it is fed to the units of the core field. 
The readout process is based on finding the optimal coefficients for the output unit. 
Regression analysis is used for minimizing the error between the input signal
and the signal produced by the output unit during the initial learning phase. 
For both stdESN and audESN, learning is implemented in an online fashion 
so that even during the prediction phase, regression analysis continues simultaneously, 
and the weights of the output unit can be updated  
periodically or at any chosen time points (for details, see appendix A).
%
%
\begin{figure}[t]
\centering
\includegraphics[width=0.9\textwidth]{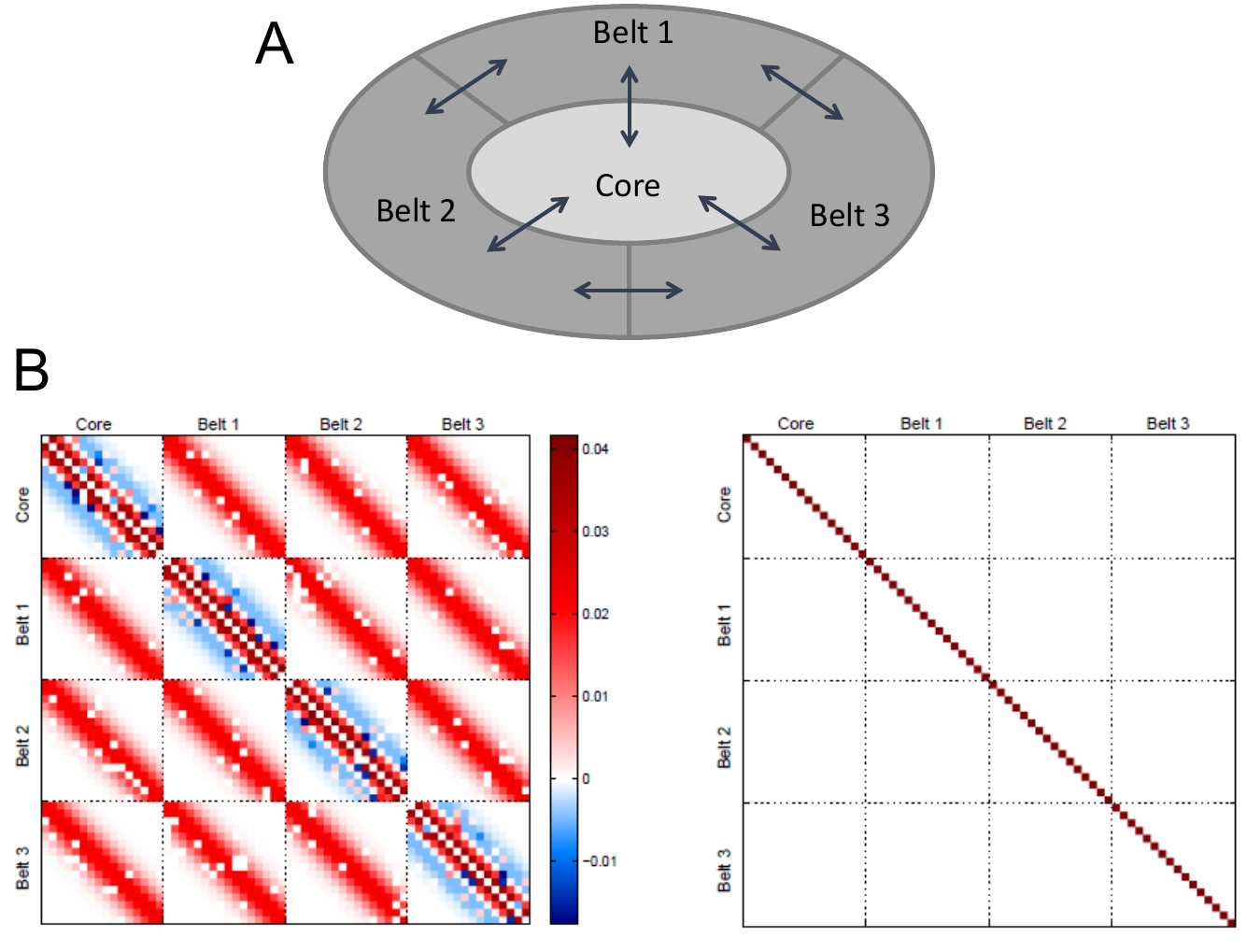}
\caption{\small
Simplified model of auditory cortex. 
(A) Schematic representation of the core-belt structure of the AC model 
used in the wave prediction algorithm. The single core field receives the input signal
which is then distributed to three surrounding belt fields. 
The arrows indicate the connections between the core and belt fields. 
(B) The connection matrices corresponding to the four fields shown in (A) are shown 
(for details, see appendix A). In the case shown here, there are $16$ columns per field. 
Each matrix element represents the strength of the connection between two columns. 
The left matrix shows the superposition of the matrices indicating excitatory-to-excitatory 
connection ($W_{ee}$, red) and excitatory-to-inhibitory connections ($W_{ie}$, blue). 
The matrix on the right shows the two (overlaid) diagonal matrices representing 
inhibitory-to-inhibitory connections ($W_{ii}$) and inhibitory-to-excitatory ($W_{ei}$) connections.}
\label{fig2}
\end{figure} 
%
%
%
%
\begin{figure}
\centering
\includegraphics[width=0.9\textwidth]{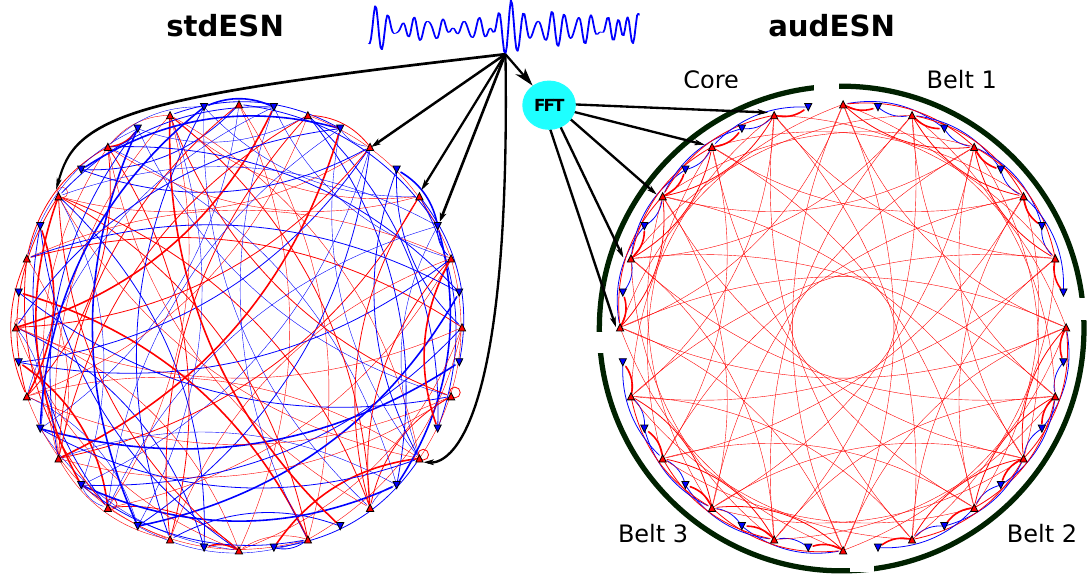}
\caption{\small
Structure of the reservoir. The blue waveform represents the incoming signal. 
On the left, the reservoir of a standard ESN (stdESN) features random connections, 
and the input feeds directly into all units (only five input connections shown). 
On the right, the simplified version of the reservoir of the auditory ESN (audESN) 
described in Fig. 2 is shown. The incoming signal to the audESN is first pre-processed 
by a Fast Fourier Transform (FFT). 
Red triangles and curves represent excitatory units and connections. 
Inhibitory units and connections are in blue. Output units are not shown.
}
\label{fig3}
\end{figure} 
%
%


\section{Prediction of ocean waves}
Real-time prediction of ocean waves is essential in marine operations for stabilizing drifters 
and other measurement equipment and for the safe and efficient operation of wind parks 
and wave energy converters \cite{8}. Early attempts to use neural network strategies 
for wave prediction were based on the stochastic theory of sea states \cite{25,26,27}. 
While these techniques have matured to the point where they can be incorporated 
into operational wave forecasts, development continues specifically with the aim of implementing 
special network structures such as gated recurrent unit (GRU) models \cite{28}. 
These methods do not provide phase-resolved forecasts, instead yielding estimates 
of average measurements of significant wave height $H_s$ and peak period $T_p$ 
with a forecasting horizon of about $24$ hours. 

Current neural network methods for phase-resolved wave predictions are based 
on deep learning \cite{10}, and as such are computationally demanding. 
Neural networks with low computational cost were used for wave-by-wave predictions 
of ocean waves in fairly deep water \cite{29}, 
but it was shown that they hold no clear advantage over traditional linear signal analysis methods 
such as auto regression techniques \cite{10}. More recent work on phase-resolved wave forecasting 
focused on testing specific network architectures such as long short-term memory (LSTM) 
networks \cite{31,32}. However, this still considers waves of relatively small steepness 
in rather deep water where bathymetric effects and wave breaking have little to no influence. 
On the other hand, in shallow water, which is the focus of this work, 
ocean waves interact with bathymetry and exhibit highly nonlinear behavior and wave breaking, 
and in such an environment, predictions are much more challenging. 
As demonstrated here, our new method enables compelling real-time operational wave-by-wave 
prediction of ocean waves in shallow water.

For testing the audESN, we used simulated wave data generated with the 
Boussinesq Ocean \& Surf Zone (BOSZ) model, a numerical modeling suite 
for free surface flow problems \cite{33}. This choice was motivated by the limitations 
of using empirical field data. This kind of data can be difficult to classify 
due to imperfect sea states, missing data due to equipment failure and other potential issues. 
Moreover, a given location may exhibit only a very narrow range of possible sea states. 
In contrast, BOSZ allows for the simulation of a broad range of sea conditions, 
and the generated data can be easily grouped into clean sea states 
which is convenient for running performance comparisons such as those in the current study. 
With BOSZ, fast and detailed computations of wave-by-wave processes over large domains 
are possible and wave breaking is modeled realistically. 
A brief description of this model is given in appendix B. 
Simulations were obtained for a range of sea states with 
significant waveheight $H_s$ between $0.5$ and $2$ m, 
wavelengths ranging from $100$ to $250$ m, and peak wave period $T_p$ from $8$ to $18$ s
(for more details, see appendix B). The BOSZ model was initialized 
with a Pierson-Moskowitz spectrum in intermediate water depth, 
and these waves are then propagated into shallow water, 
where a time series at a fixed location is extracted.

Figure 4 shows an example of how the free surface elevation traced by the stdESN (top row) 
and the audESN (bottom row) evolves across a time window of about $37$ minutes. 
To provide the surface elevation to be predicted, BOSZ simulations (black curve) 
were performed using a wave height of $2$ m and a peak wave period of $12$ s. 
In each case, the input signal to the ESN is the surface elevation at the same location 
where it is predicted. After the initial $900$ s training phase, 
training was updated online at every other wave trough. 
The difference in prediction performance is clearly visible in the outputs (red curves) 
of stdESN and audESN. This is particularly evident in the zoomed-in representation of a shorter, 
$50$-s long time window, where the prediction accuracy of stdESN falls significantly below that of audESN, 
especially for larger free surface elevations. 

ESNs with $16$ different reservoir types were tested, comprising stdESN, audESN, 
and $14$ intermediate versions (Fig. 5A). 
These reservoir types are represented by a four-digit code where each digit expresses 
a binary construction choice. The first digit indicates the neuron model 
(presynaptic: 0; postsynaptic: 1). The second digit refers to the architecture, 
i.e. the type of connectivity (random: 0; tonotopic: 1). 
The third digit expresses whether the input was in the time domain, i.e. fed directly to the
reservoir (digit 0) or in the frequency domain, i.e. FFT pre-processed (digit 1). 
The fourth digit indicates the size of the reservoir (small: 0; large: 1). 
Hence, the two default stdESNs are represented by ‘0000’ ($32$ units) and ‘0001’ ($128$ units). 
These had the presynaptic neuron model, random connectivity between the nodes, 
and computations performed in the time domain. The two default audESNs 
are represented by ‘1110’ (small/flat) and ‘1111’ (large/hierarchical). 
These utilize the postsynaptic neuron model, organization into tonotopic fields, 
and transformation of the signal into the frequency domain. As one moves from left to right in Fig. 5A, 
there is a gradual transition from stdESN to audESN. 

In each test, the initial training period was $900$ s, and the training was subsequently updated online at every 
other wave trough (see appendix A). Prediction performance was quantified as the Root Mean Square (RMS)
 error over $1.5$ h of wave data. Each column of Fig. 5A shows the prediction performance 
in five selected sea states indicated by the ticks on the abscissa and characterized 
by the significant wave height $H_s$ and the peak period $T_p$. From left to right, 
the five sea states are: 
$[0.5 \mathrm{m},  8 \mathrm{s}]$, 
$[  1 \mathrm{m},  8 \mathrm{s}]$, 
$[  1 \mathrm{m}, 10 \mathrm{s}]$, 
$[  1 \mathrm{m}, 12 \mathrm{s}]$, and 
$[  2 \mathrm{m}, 10 \mathrm{s}]$. 
For each sea state, $125$ RMS values are given. Each RMS value is for a network with a unique combination 
of the three network parameters that were varied: 
leak rate $\alpha$, spectral radius $\rho$  and maximum bias $\beta$ (for details, see appendix A). 
Each of these was independently varied in five steps $(0.1, 0.3, 0.5, 0.7, 0.9)$, 
resulting in $125$ combinations.

The following observations can be noted: 
First, by far the lowest RMS values were obtained with the small (’1110’) and large (’1111’) audESN; 
these RMS values were much smaller than those for the corresponding stdESNs (‘0000’ and ‘0001’). 
The best performing network was the large audESN (’1111’), with the hierarchical structure of four tonotopic fields. 
The small audESN (’1110’) with only one single field performed slightly worse. 
Second, the use of the postsynaptic neuron model of audESN (first digit = 1) 
consistently leads to lower minimum RMS values than the presynaptic model of stdESN (first digit = 0). 
Third, within each network type, the larger the waves of the sea state 
(i.e. as one moves from left to right within each column), the larger the RMS values. 
In almost all cases, the largest sea state [2 m, 10 s] leads to a very large scatter of RMS values. 
The exception to this is the performance of the small and large audESN 
where the scatter remains small for all sea states.

These observations are displayed in more detail in Fig. 5B 
which shows the performance data for the five sea states for stdESN (‘0001’, open symbols) 
and audESN (‘1111’, filled symbols). The median RMS of audESN is smaller compared to the median RMS of stdESN 
for all sea states, although the differences between audESN and stdESN are small for small sea states. 
However, this difference increases with increasing sea severity 
(higher significant wave height) and reaches a value of about two 
for the largest sea state examined. Importantly, the performance of audESN 
shows a much larger robustness against parameter and sea-state choice, 
as expressed in the minute confidence intervals (CIs) compared to the large and widely distributed 
(CIs) obtained for stdESN, especially for the largest sea state. 
In general, the audESN outperformed all other configurations.

%
%
\begin{figure}[t]
\centering
\includegraphics[width=0.9\textwidth]{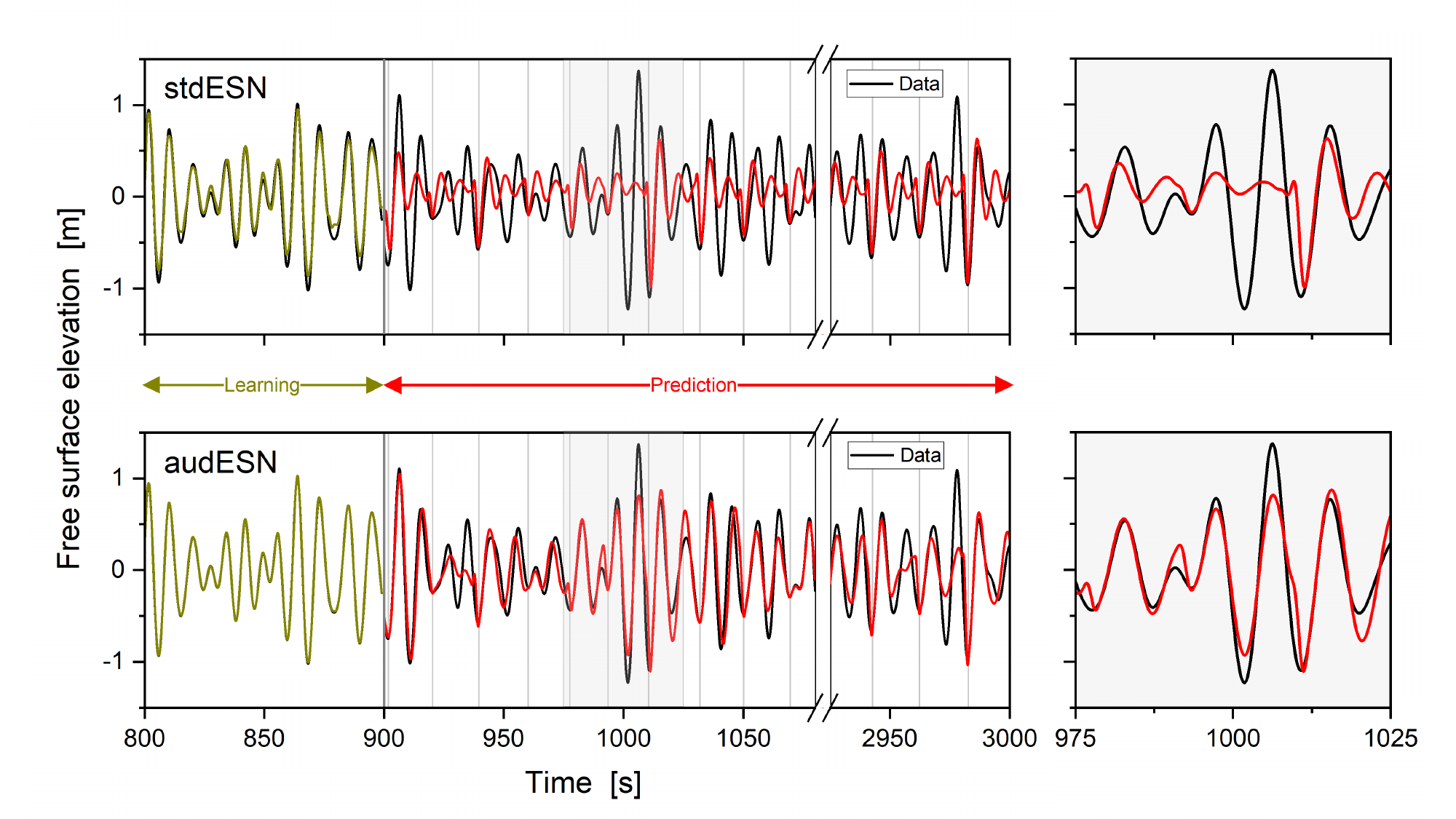}
\caption{\small
Prediction performance of the networks. This example shows how stdESN (top row) 
and audESN (bottom row) track the wave data (black curve) during training ($< 900$ s, green curve) 
and prediction ($> 900$ s, red curve). The network training was updated online 
at every other wave trough, as indicated by the vertical lines.  
Clearly, audESN provides much better predictions than stdESN 
which is evident in the two enlarged time windows shown on the right.}
\label{fig4}
\end{figure} 
%
%
%
%
\begin{figure}
\centering
\includegraphics[width=0.9\textwidth]{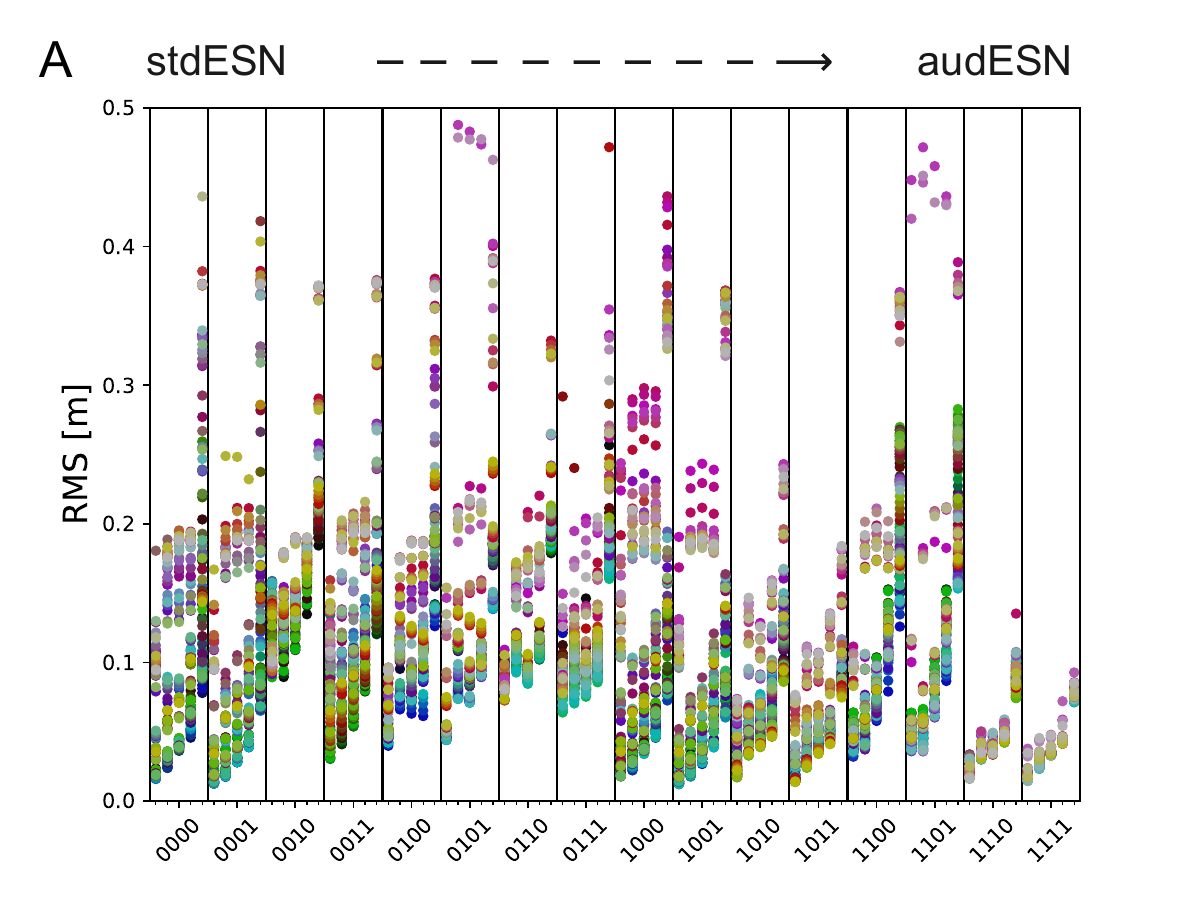}\\
\includegraphics[width=0.6\textwidth]{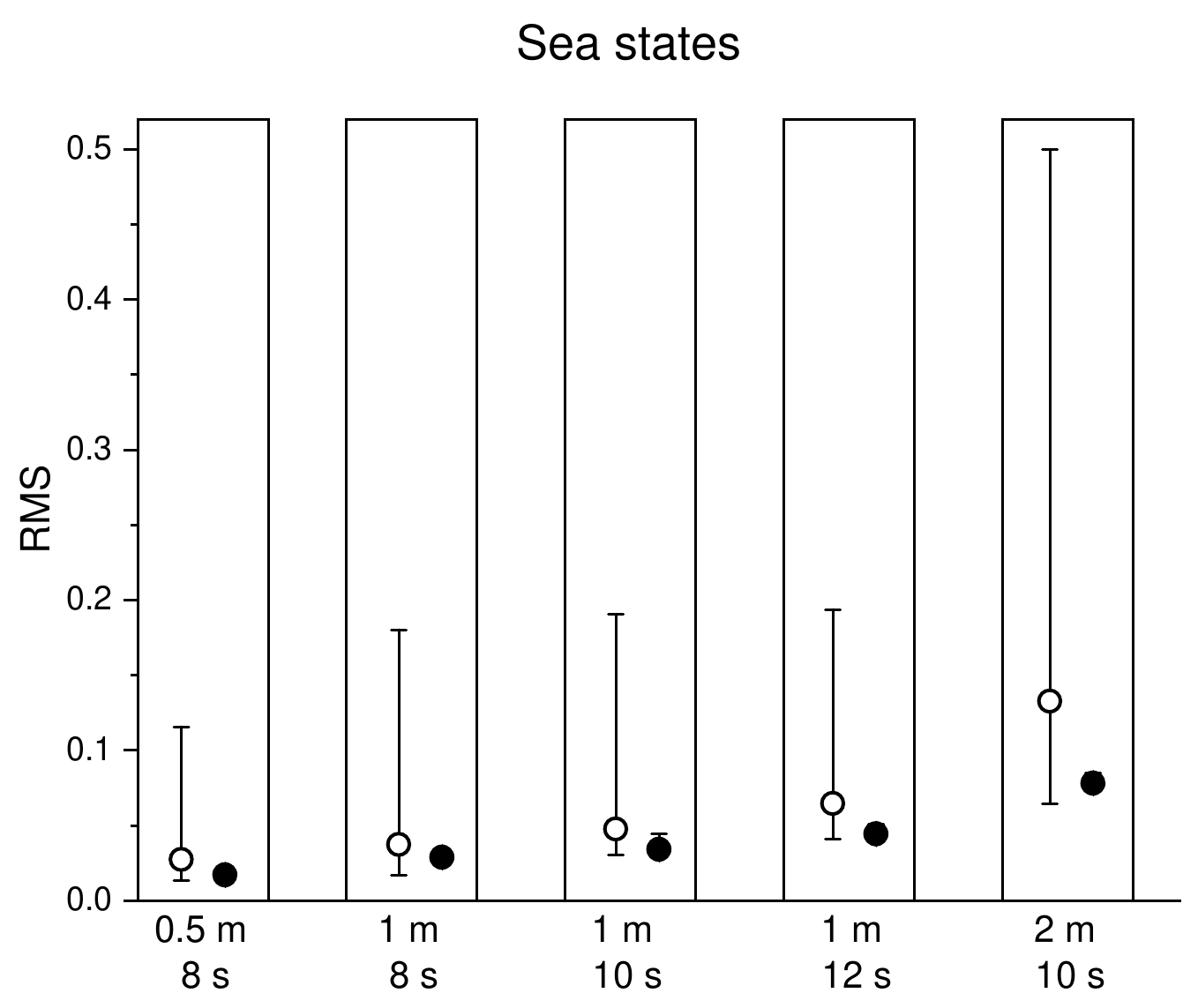}
\caption{\small
Prediction performance. (A) Example of the prediction performance of $16$ networks. 
The figure shows, for each network type, the RMS error between simulated and predicted waves 
over $1.5$ h for five sea states whose significant wave height (first row) 
and peak wave period (second row) are given at the bottom of the figure. For each sea state, 
tests were carried out for 125 combinations of the three network parameters: leak rate $\alpha$, 
spectral radius $\rho$, and maximum bias $\beta$ (for details, see appendix A). 
The four-digit codes on the abscissa denote the four binary network construction choices: 
neuron model, architecture, input domain, and network size (see section 3 and appendix A). 
(B) The median RMS errors of stdESN (‘0001’, open circles) and audESN (‘1111’, filled circles) 
with the corresponding confidence intervals (CIs) are shown.  
Among all sea states, the lowest RMS errors are achieved with audESN. 
Importantly, audESN also shows the best performance stability against variations 
in sea state and parameter choice. Note that the CIs for all five sea states are 
so small that they are covered by the size of the median symbol.}
\label{fig5}
\end{figure} 
%
%

\section{Conclusions}
%
%
While the development of artificial neural networks was originally inspired by attempts to model 
neuronal computations \cite{1,2}, there have not been many studies where models of the brain 
or other biological structures are utilized as a blueprint for the organization of an ANN \cite{30}. 
As shown here for shallow water conditions, using a network based on the auditory cortex architecture 
and the preprocessing of the signal to represent its frequency structure gives rise to a powerful, 
computationally light tool for forecasting time series. This configuration mimics 
a biological neural network which is already optimized by natural selection to  
deal with single point measurements of time series data, that is, sound waves arriving at the ear. 
The resulting increase in operational efficiency means that the size of the network 
can be reduced drastically. For the predictions of ocean waves shown in Figs. 4 and 5, 
a network with 128 units was sufficient to provide an acceptable prediction on a forecasting 
horizon of two wave periods. This contrasts with other ANN methods, such as LSTM, 
used for generating phase-resolving predictions of ocean waves \cite{31,32}. 
In these computationally heavier networks, the number of units tends to be an order of magnitude larger, 
and all weights in the model are optimized, as opposed to just the output weights, 
as in the ESN-based approach. We note that the current ESN approach 
is perfectly suited for demonstrating the superiority of biological design. 
The fixed nature of the reservoir means that its organization 
- tonotopic connectivity over multiple cortical fields in this case - 
is retained throughout the training process. This allows for a straightforward comparison 
of different network structures in terms of performance. This would be more complicated 
with ANN methods where all the weights are optimized and where any initial setup 
of the internal structure is unlikely to survive the training process.

The relatively small size of the reservoir enables updating of the output weights almost instantaneously, 
in microseconds, through real-time simultaneous learning even during the prediction phase. 
In effect, audESN enables continuous operation without any need to retrain the network. 
In fact, this special feature of constant readjustment of the network weights 
even in ongoing operation leads to optimal wave prediction under virtually any realistic wave conditions. 
In the present work, the system was tested for individual sea states, that is, 
under stationary conditions, but the continuous learning and online updating of the weights 
allow operations even under changing sea conditions.
This appears to be a particularly attractive feature for the implementation 
of the framework for real-time wave prediction, especially near the shore. 
Indeed, to the best of our knowledge, no existing data-based wave prediction method has been tested 
in the shallow-water conditions of the nearshore zone, where bathymetric effects, 
nonlinear steepening, and wave breaking combine to create a highly complex wave signal. 
For this reason, it is difficult to make direct comparisons on performance with previous 
ANN approaches that have focused on predictions in deep-water conditions \cite{31,32}.

Due to the good performance in shallow-water conditions and the robustness 
with regard to both reservoir parameters and sea states, there is no need for ensemble forecasts. 
The network can be trained a priori using data generated from ocean or nearshore models 
such as BOSZ. The lightweight nature of the network allows operation on embedded and even handheld devices
can potentially be implemented directly in hardware, and is thus ideal for deployment at sea.

The proposed analogy between ESNs and the auditory system should not be stretched too far. 
First, the wave periods of ocean waves are orders of magnitude longer than those of perceptible sound waves. 
The current ESN is not a model of the auditory cortex as such. Rather, it copies the general principle of transducing 
the one-dimensional input into representations on multiple frequency maps. 
Second, the current results do not imply that the auditory cortex actually operates 
like an echo-state network, with some output neurons tracing predicted activity 
(although this has been suggested as a possible mechanism of predictive coding in the auditory system \cite{34}). 
However, the analogy does offer possibilities of further integration of neuroscience into the ESN framework. 
In short-term plasticity, synaptic strengths in auditory cortex change dynamically 
according to pre-synaptic activity \cite{35}. 
This plasticity is an order of magnitude slower than the membrane and spiking dynamics, 
and it offers a possible memory mechanism for representing sounds over longer intervals \cite{18,19}. 
Further, amplitude modulations of sound signals are also represented in auditory cortex so that individual neurons are
tuned to specific amplitude modulation frequencies \cite{36}. 
Adding these features to the ESN reservoir design might allow 
for further improvements in performance in terms of lengthening the effective prediction window.  

In the audESN approach, changes in the frequency structure of the input signal 
lead to maximal changes in the reservoir state. This is due to frequency being represented spatially 
in multiple tonotopically-organized fields and because the connections between the fields 
are partly random so that each field represents the signal in a unique way. Although not shown formally, 
the effect is to maximize the entropy of the reservoir states as well as the mutual information 
between the spectral structure of the input and the spatial constellation of activity in the reservoir. 
This approach leads to better ESN predictive ability, and we suspect that it also describes 
the relationship between sound structure and the activity of auditory cortex. 
This might explain why there are so many tonotopic, 
hierarchically organized fields in auditory cortex - one of the mysteries of auditory neuroscience.   \\\\

\textbf{Acknowledgements.}
We thank Jussi Åkerberg and Christopher Ridgewell at AW-Energy, Finland, for putting together and administering the consortium of researchers, which made this study possible. PM is especially grateful to CR for the discussions that germinated the idea of mixing auditory neuroscience, machine learning, and ocean wave prediction. We thank Aida Hajizadeh and Robert Klank for their help in preparing the figures, and Peter Heil for his constructive criticism of the manuscript. This work received funding from the European Union's Horizon 2020 research and innovation programme (grant agreement no. 763959), and from the European Regional Development Fund (ERDF; project ZS/2018/11/96066). HK acknowledges funding from Bergen Universitetsfond. VR acknowledges financial support from the I-SITE program Energy \& Environment Solutions (E2S), the Communauté d'Agglomération Pays Basque (CAPB) and the Communauté Région Nouvelle Aquitaine (CRNA) for the chair position HPC-Waves.  \\\\





\section*{Appendix A: Reservoir computing}
\subsection*{A.1.  Model of auditory system}
\label{sec:acmodel}
The time series prediction method developed in this paper is based on replacing the
randomly connected reservoir of an Echo State Network by a structured reservoir constructed
in accordance with our current understanding of the auditory cortex.
The auditory cortex is a complex biological system which has been under study for some time.
In a series of papers, May and colleagues developed a computational model of the cortex
which is able to reproduce many phenomena observed in studies of the
auditory system \cite{18,19}.

The model is based on the idiosyncratic architecture of the auditory system of primates \cite{16}. 
The basic computational unit of this AC model is the canonical microcircuit, originally proposed 
on the basis of results from the visual cortex of cats. It consists of interacting
excitatory and inhibitory cell populations, which receive afferent 
feedforward excitation from the thalamus. Although
this is a vast simplification of real cortical columns, which have
an intricate, layered structure, the model captures the basic observation
whereby all neurons within a column respond in the same way to frequency
input to the column, that is, they exhibit similar tuning. The weak
thalamocortical input to the excitatory population is essentially
amplified by strong recurrent excitation, which itself is damped
by local feedback inhibition.

A central element of the original AC model 
is short-term synaptic depression (STSD),
which is realized by modifying the excitatory connections by time-dependent
depression terms that depend on presynaptic firing.
Synaptic depression has a lifetime of seconds 
and can be considered to be a local form of memory.
However, when it is introduced to a complex network,
STSD leads to various adaptation phenomena as well as to emergent
properties such as individual cortical units showing selectivity to
the temporal structure of sound. 
The model explains a wide range of phenomena observed in invasive (electrophysiological) and non-invasive (magnetoencephalographic) experiments.

\subsection*{A.2. The standard ESN framework}
\label{subsec:stdESN}
In a classical ESN framework \cite{11}, 
the time series representing the incoming wave
is fed as input to a neural network---the so-called dynamic reservoir---with
fixed structure and weights, and the units of the network are connected
to an output unit. In most applications, the reservoir is a network of randomly connected units of a given size $N$.
We refer to this network as standard ESN (stdESN). During training, only the
output weights are changed through linear regression,
which makes ESNs computationally very efficient. 
After training, the output signal resembles the training signal.
Once this
has been achieved, the system can be used for prediction
by switching off the input signal and replacing it with feedback from
the output. 

The state equation of an stdESN is given by
\begin{equation}
\label{eq:stdESN}
\dot{x}_{j}(t)=-\alpha x_{j}(t)+\tanh\Big(\left[\rho\mathbf{W}\mathbf{x}(t)\right]_{j}+\beta_{j} + \left[\mathbf{D} \mathbf{s}(t)\right]_{j} \Big),
\end{equation}

\noindent
where $x_{j}(t)$ ($j=[1,2,...,N]$) represents the dynamical behavior of the individual
neurons in the reservoir with the parameters
leak rate $\alpha$, 
spectral radius $\rho$, and maximum bias $\beta$, whereby biases $\beta_j$ are randomly chosen from a uniform distribution $\beta_j \in\mathit{\mathfrak{U}(-\beta,\beta)}$.
The  random connectivity of the neurons of the 
stdESN is expressed by the matrix $\mathbf{W}_{N \times N} \in \mathit{\mathfrak{U}}(-1,1)$ 
with spectral scaling 
$\mathbf{W}\rightarrow{\mathbf{W}}/{{\textrm{maxeig}}(\mathbf{W})}$. 
The stdESN is driven by an $M$-dimensional  stimulus 
$\mathbf{s}_{M \times 1} (t)$ multiplied by a random input matrix $\mathbf{D}_{N \times M} \in\mathfrak{U}(-1,1)$. 
Matrix multiplications are given by $[\mathbf{W}\mathbf{x}{(t)}]_{j} = \sum_{l=1}^{N} W_{jl} x_{l}(t)$ and
$[\mathbf{D}\mathbf{s}(t)]_{j} = \sum_{l=1}^{M} D_{jl} s_{l}(t)$. 
This ESN features what we term a 
\textit{presynaptic}
nonlinearity, where the nonlinear $\tanh$ function is applied to all terms
except the leak term $\alpha x_j(t)$.

\subsection*{A.3. The auditory system ESN framework}
\label{subsec:audESN}

\noindent
\subsubsection*{A.3.1. Overview}

\noindent
Our innovation lies in designing the network to mimic the main aspects of the auditory system.
To achieve this, the incoming data are pre-processed via spectral 
analysis using real-time Fast Fourier Transform (FFT). The output
is then fed into a set of spatial frequency maps which are tonotopically organized. 
The units in each tonotopic map are connected to each other with a mixture of
excitation and lateral inhibition, in roughly the manner one finds
in the auditory system. In contrast to the full AC model
used in neuroscientific research \cite{17,19} 
(see e.g. Fig.\,1 of the main text), we simplified the 
auditory system structure used for audESN such that the reservoir 
comprises only four tonotopic maps, which are organized in a 
core-belt structure. 
Specifically, the central core field receives the pre-processed
input and distributes the signal to three surrounding belt fields
which are connected in a ring around the core, as displayed in 
Fig.\,2A of the main text.  \\

\noindent
\subsubsection*{A.3.2. Auditory cortex dynamics}

\noindent
In the AC model, the dynamics are described by two coupled
differential equations for the state variables of the excitatory neurons $\mathbf{u}(t)$ (indexed by subscript `e') and the inhibitory neurons
$\mathbf{v}(t)$ (indexed by subscript `i') \cite{21}. The equations are written in the form
\begin{equation}
\begin{array}{c}
\tau_{\textrm{m}}\dot{\mathbf{u}}(t)=-\mathbf{u}(t)+\mathbf{W}_{\textrm{ee}}g[\mathbf{u}(t)]-\mathbf{W}_{\textrm{ei}}g[\mathbf{v}(t)]+\mathbf{I}_{\textrm{e}}{(t,f)}\\

\tau_{\textrm{m}}\dot{\mathbf{v}}(t)=-\mathbf{v}(t)+\mathbf{W}_{\textrm{ie}}g[\mathbf{u}(t)]-\mathbf{W}_{\textrm{ii}}g[\mathbf{v}(t)]+\mathbf{I}_{\textrm{i}}{(t,f)}. 
\end{array}
\label{eq:uv}
\end{equation}

\noindent
Here, $\tau_{\textrm{m}}$ is the membrane time constant. The connections between the neurons are expressed by four weight matrices describing
the excitatory-to-excitatory ($\mathbf{W}_{\textrm{ee}}$) and excitatory-to-inhibitory
($\mathbf{W}_{\textrm{ie}}$) connections, respectively (Fig.\,2B, left matrix)
as well as the inhibitory-to-inhibitory ($\mathbf{W}_{\textrm{ii}}$) and 
inhibitory-to-excitatory ($\mathbf{W}_{\textrm{ei}}$) connections 
(Fig.\,2B, right matrix).
In terms of modeling the auditory cortex, each neuron is representative of a population of neurons within a cortical column.
Excitatory neurons make connections locally, laterally within a field, as well as across fields.
Inhibitory neurons make connections locally only, within the column, and therefore $\mathbf{W}_{\textrm{ii}}$ and $\mathbf{W}_{\textrm{ei}}$ are diagonal.
The firing rates $g[\mathbf{u}(t)]$ 
and $g[\mathbf{v}(t)]$ for the
excitatory and inhibitory neurons are sigmoid functions 
$g[x] = \tanh(x)$. 
The terms $\mathbf{I}_{\textrm{e}}{(t,f)}$ and 
$\mathbf{I}_{\textrm{i}}{(t,f)}$ 
are time- and frequency-dependent afferent inputs and target only the core area. 
Here we note that the size 
$F$
of each field, 
i.e.\ the number of columns per field
in the network, equals the number of frequencies (the window length used for FFT).
The size $N$ of the network in terms of neurons is determined by the number of 
fields (four), the number of
columns per field ($F$), 
and the number of
neurons per column (two, one excitatory and one inhibitory),
so that $N = 2*4*F$.

To enable a systematic comparison of audESN with stdESN as shown in Eq. \eqref{eq:stdESN}, we 
multiplied the connectivity matrices $\mathbf{W}_{\textrm{ee}}$, $\mathbf{W}_{\textrm{ie}}$, $\mathbf{W}_{\textrm{ei}}$ and $\mathbf{W}_{\textrm{ii}}$
with a parameter $\rho$ corresponding to the spectral radius ($\mathbf{W}_{\textrm{xx}} \rightarrow \rho \mathbf{W}_{\textrm{xx}}$),
and added a parameter corresponding to the maximum bias $\beta$, that is, the individual random biases are
$\boldsymbol{\beta}_{\textrm{e}} \in \mathit{\mathfrak{U}(-\tau_{\textrm{m}} \beta ,\tau_{\textrm{m}} \beta)}$ 
and $\boldsymbol{\beta}_{\textrm{i}} \in \mathit{\mathfrak{U}(-\tau_{\textrm{m}} \beta ,\tau_{\textrm{m}} \beta)}$.
The input matrices $\mathbf{D}_{\textrm{e}}$ and $\mathbf{D}_{\textrm{i}}$ are tonotopically organized, that is,  diagonal.  
Hence, Eq. \eqref{eq:uv} can be transformed into
\begin{equation}
\begin{array}{c}
\tau_{\textrm{m}} \dot{\mathbf{u}}(t)= -\mathbf{u}(t)+\rho\mathbf{W}_{\textrm{ee}}\tanh(\mathbf{u}(t))
                                        -\rho \mathbf{W}_{\textrm{ei}}\tanh(\mathbf{v}(t))
                                        +\boldsymbol{\beta}_{\textrm{e}}+\mathbf{D}_{\textrm{e}}\mathbf{I}_{\textrm{e}}(f,t)  \\

 \tau_{\textrm{m}} \dot{\mathbf{v}}(t)= -\mathbf{v}(t)+ \rho \mathbf{W}_{\textrm{ie}}\tanh(\mathbf{u}(t))
                                        -\rho \mathbf{W}_{\textrm{ii}}\tanh(\mathbf{v}(t))
                                         +\boldsymbol{\beta}_{\textrm{i}}+\mathbf{D}_{\textrm{i}}\mathbf{I}_{\textrm{i}}(f,t). 
\end{array}
\label{eq:uv2}
\end{equation}

\noindent
Dividing Eq. \eqref{eq:uv2} by $\tau_{\textrm{m}}$ and formally combining  the equations of the two state variables $\mathbf{u}(t)$ 
and $\mathbf{v}(t)$ according to
\begin{equation}
\mathbf{x}(t)=\left[\begin{array}{c} \mathbf{u}(t) \\ \mathbf{v}(t)\end{array}\right]_{N \times 1}
\end{equation}
\noindent
then leads to
\begin{equation}
\dot{x}_{j}(t)=-\alpha x_{j}(t)+\left[\rho\mathbf{W}\tanh(\mathbf{x}(t))\right]_{j}+\beta_{j}+\left[\mathbf{D}\mathbf{s}(t)\right]_{j},
\label{eq:audESN}
\end{equation}
\noindent
where $\alpha = 1/\tau_{\textrm{m}}$ is the leak rate. The connectivity matrices ($\mathbf{W}_{\textrm{ee}}$, $\mathbf{W}_{\textrm{ie}}$, 
$\mathbf{W}_{\textrm{ei}}$, $\mathbf{W}_{\textrm{ii}}$) are combined into
\begin{equation}
\mathbf{W}=\frac{1}{\tau_{\textrm{m}}}\left[\begin{array}{cc}
\mathbf{W}_{\textrm{ee}} & -\mathbf{W}_{\textrm{ei}}\\
\mathbf{W}_{\textrm{ie}} & -\mathbf{W}_{\textrm{ii}}
\end{array}\right]_{N \times N}
\end{equation} 
and spectral scaling is applied in accordance with stdESN, that is, $\mathbf{W}\rightarrow{\mathbf{W}}/{{\textrm{maxeig}}(\mathbf{W})}$. The bias terms are given by
\begin{equation}
\boldsymbol{\beta}=\frac{1}{\tau_{m}} \left[\begin{array}{cc} \boldsymbol{\beta}_{\textrm{e}} \\ \boldsymbol{\beta}_{\textrm{i}} \end{array} \right]_{N \times 1}
\end{equation}
and the input matrices are combined into a single matrix $\mathbf{D}$.
Since the signal is decomposed into $F$ frequencies (with real and imaginary parts) and feeds only into the core area, $\mathbf{D}$ has the dimensions of the input $M = 2*F$ times the network size $N$:
\begin{equation} 
\mathbf{D}^T=\frac{1}{\tau_{m}}\left[\begin{array}{cccccccc}
\mathbf{D}_{\textrm{e}} & \mathbf{0} & \mathbf{0} & \mathbf{0} & \mathbf{0} & \mathbf{0} & \mathbf{0} & \mathbf{0}\\
\mathbf{0} & \mathbf{D}_{\textrm{i}} & \mathbf{0} & \mathbf{0} & \mathbf{0} & \mathbf{0} & \mathbf{0} & \mathbf{0} 
\end{array}\right]_{M \times N},
\end{equation}
where the transpose of $\mathbf{D}$ is given for readability.
The multidimensional input signal is
\begin{equation}
\mathbf{s}(t)=\left[\begin{array}{cc} \mathbf{\mathbf{I}}_{\textrm{e}}(t,f) \\ \mathbf{I}_{\textrm{i}}(t,f) \end{array} \right]_{M \times 1}.
\end{equation}

\noindent
Eq. \eqref{eq:audESN} is equivalent to Eq. \eqref{eq:stdESN}, however with one major difference: 
unlike in Eq. \eqref{eq:stdESN}, the nonlinear $\tanh$ function is applied only to the state variable $\mathbf{x}(t)$.
Thus, we say that Eq. \eqref{eq:audESN} realises a neuron model with \textit{postsynaptic} nonlinearity.

\subsubsection*{A.3.3. Network properties}
\label{subsubsec:netw-props}
Both stdESN and audESN are characterized by the following four network properties: 

\vskip 0.1in

\noindent
1) \textit{The neuron model}. This property indicates whether it is a presynaptic 
(stdESN) or a postsynaptic (audESN) model. In the presynaptic case, 
the nonlinear (sigmoidal) firing rate function is applied after the linear summation of
all units connected to a given unit has been performed. The audESN utilizes a 
postsynaptic neuron model, where the nonlinear sigmoidal function is applied first, 
prior to the multiplication with the weight matrices. 

\vskip 0.1in

\noindent
2) \textit{Connectivity between neurons}. In stdESN, there is no predefined organization 
of the network; the neurons are connected randomly. In audESN, the 
connections between neurons follow a well-defined anatomically-inspired structure. 
The audESN consists of excitatory and inhibitory connections, and the neurons 
are organized tonotopically. Further, within each field, neighboring neurons are more strongly connected with each other
than those that are farther away from each other. 

\vskip 0.1in

\noindent
3) \textit{Time or frequency domain}. 
The working domain for stdESN is the time domain, so the input to the reservoir
is just the original (not Fourier-transformed) signal. In audESN, the input signal
first undergoes a windowed Fourier transform, and the transformed signal is 
subsequently distributed tonotopically to the respective columns.

\vskip 0.1in

\noindent
4) \textit{The size of the network}. 
We used small ($N=2*1*16$) and large ($N=2*4*16$) networks for both stdESN and audESN. 
The small audESN has only a single tonotopically organised field whereas the large audESN has four, as explained above.
These two therefore differ not only in size but also with regard to the absence or 
presence of a hierarchical organisation.

\vskip 0.1in

The above four properties can be realised in $2^{4}=16$ network types. We tested all these network types for their ability to forecast ocean wave signals.

\subsubsection*{A.3.4. The prediction algorithm}
%

\noindent
{\em General description of the algorithm}
%

\noindent
For stdESN, the input representing the incoming waves is fed as a time-domain signal to all the neurons of the reservoir. 
In contrast, to mimic the processing of an auditory signal in the cochlea of the
inner ear, the algorithm for audESN processes the input
signal $s(t)$ using short-time Fourier
transform in real time. As the ocean waves are low-frequency
oscillations where phase plays a crucial role, we use the
real part of the FFT as input to the excitatory neurons of the AC
network, while the imaginary part acts as input to the inhibitory neurons:
\begin{equation}
[\Re(FFT(s(t))),\Im(FFT(s(t))] \rightarrow \mathbf{s}_{M \times 1}(t). \label{eq:Input}
\end{equation}

The input drives the reservoir, whose design represents the combination of the four network properties
explained in Sec. A.3.1. 
The readout process is based on finding the optimal
weights $\mathbf{R}$ by which 
the neurons are connected to the
output unit (the so-called readout neuron). 
Regression analysis is used for minimizing the root-mean-square (RMS) error between
the input signal and the signal produced by the output unit, as given by Eq. \eqref{eq:stdESN} 
and Eq. \eqref{eq:audESN}.
Note that in the frequency domain, the one-dimensional ocean wave height signal $s(t)$ 
is transformed into the multidimensional signal $\mathbf{s}_{M \times 1}(t)$ according to Eq. \eqref{eq:Input}.

Once the optimal ESN output weights have been determined, a prediction is made by cutting off 
the input to the ESN and replacing it with the output signal. The state of the reservoir continues 
to generate an output signal $\widetilde{s}(t)=\mathbf{R}\mathbf{x}(t)$. 
Note that in cases where FFT is applied to the input signal $s(t)$, $x(t)$ 
also represents the signal in the frequency domain.
To generate the output signal, 
the inverse FFT from the frequency domain to the time domain is applied: 
$\text{IFFT}(\widetilde{\mathbf{s}}(t)) \rightarrow \widetilde{s}(t)$.
The output weights can be updated in real time with online regression,
as described in the following section. \\

%

\noindent
\textit{Online regression}

\noindent
For efficient and fast operation, the weights between the nodes of
the network have to be updated continuously. Simple regression
analysis requires the computation of the inverse of a matrix, which,
however, can be evaded by using the following approach. Assuming the
dynamics matrix will be of the form $\mathbf{X}_{t_{j+1}}=[\mathbf{x}_{t_{j}},\mathbf{x}_{t_{j+1}}]$,
then the online update of the regression weights $\mathbf{R}$ with
new input data points $\mathbf{s}_{t_{i+1}}$ are given by

\begin{equation}
\mathbf{R}{}_{t_{j+1}}=(\mathbf{X}_{t_{j+1}}^{T}\mathbf{X}_{t_{j+1}}+r\mathbf{I})^{-1}\mathbf{X}_{t_{j+1}}^{T}\mathbf{s}_{t_{j+1}}
\end{equation}

\begin{equation}
\mathbf{X}_{t_{j+1}}^{T}\mathbf{X}_{t_{j+1}}+r\mathbf{I}=\mathbf{X}_{t_{j}}^{T}\mathbf{X}_{t_{j}}+\mathbf{x}_{t_{j+1}}\mathbf{x}_{t_{j+1}}^{T}+r\mathbf{s}_{t_{j+1}}.
\end{equation}

\noindent
Defining the matrix $\mathbf{P}_{t_{i}}=(\mathbf{X}_{t_{i}}^{T}\mathbf{X}_{t_{i}}+r\mathbf{I})^{-1}$
we get
\begin{equation}
\mathbf{P}_{t_{j+1}}^{-1}=\mathbf{P}_{t_{j}}^{-1}+\mathbf{x}_{t_{j+1}}\mathbf{x}_{t_{j+1}}^{T}.
\end{equation}
Using matrix inversion $(\mathbf{A}+\mathbf{B}\mathbf{C}\mathbf{D})^{-1}=\mathbf{A}^{-1}-\mathbf{A}^{-1}\mathbf{B}(\mathbf{C}^{-1}+\mathbf{D}\mathbf{A}^{-1}\mathbf{B})^{-1}\mathbf{D}\mathbf{A}^{-1}$
with $\mathbf{A}=\mathbf{P}_{t_{j}}^{-1}$ , $\mathbf{B}=\mathbf{x}_{t_{j+1}}$ , $\mathbf{C=}\mathbf{I}$  
and $\mathbf{D}=\mathbf{x}_{t_{j+1}}^{T}$,
the update can be expressed as
\begin{equation}
\mathbf{P}_{t_{j+1}}=\mathbf{P}_{t_{j}}(\mathbf{I}-\frac{\mathbf{x}_{t_{j+1}}\mathbf{x}_{t_{j+1}}^{T}\mathbf{P}_{t_{j}}}{1+\mathbf{x}_{t_{j+1}}^{T}\mathbf{P}_{t_{j}}\mathbf{x}_{t_{j+1}}}),
\end{equation}
\noindent
where $1+\mathbf{x}_{t_{j+1}}^{T}\mathbf{P}_{t_{j}}\mathbf{x}_{t_{j+1}}$
is a scalar. One can also simultaneously downgrade previous data points
to have an $n$-windowed regression, that is
\begin{equation}
\mathbf{P}_{t_{j-n}}=\mathbf{P}_{t_{j}}(\mathbf{I}+\frac{\mathbf{x}_{t_{j-n}}\mathbf{x}_{t_{j-n}}^{T}\mathbf{P}_{t_{j}}}{1-\mathbf{x}_{t_{j-n}}^{T}\mathbf{P}_{t_{j}}\mathbf{x}_{t_{j-n}}}).
\end{equation}
When the next data point $\mathbf{y}_{t_{j+1}}$
arrives, the updated weights $\mathbf{R}_{t_{j+1}}$ are
\begin{equation}
\mathbf{R}_{t_{j+1}}=\mathbf{P}_{t_{j+1}}\mathbf{X}_{t_{j+1}}^{T}\mathbf{s}_{t_{j+1}}.
\end{equation}
\noindent
This procedure avoids the explicit, time consuming calculation of the matrix inversion of
$\mathbf{P}_{t_{j+1}}$. \\

\noindent
\textit{Online spectral analysis}

For efficient updating of the prediction in the frequency domain, 
a sliding window of $F$ samples moving across the time series is applied. 
At each update, one does not need to compute the whole FFT across the window. 
Instead, to compute the FFT at $t = t_{i+1}$, one can utilize the FFT computation 
from the previous time step $t = t_i$.  
\begin{equation}
\text{FFT}(\mathbf{s}_{[t_{i-n+1},t_{i+1}]})_k=-i(\text{FFT}(\mathbf{s}_{[t_{i-n},t_{i}]})_k+(s_{t+1}-s_{t-n}))e^{i2\pi f_{k}}
\end{equation}
for a given frequency $f_k$.
This procedure, combined with the above online regression method allows for rapid, 
on-the-fly updating of the system while it is in the prediction phase.

\subsubsection*{A.3.5. Wave-by-wave predictions}
We systematically investigated the prediction performance of 16 different types of ESN, 
the comprised stdESN, audESN, as well as $14$ intermediate network types, 
as explained in Sec.\, A.3.1.
As data, we used the simulated waveforms computed with the BOSZ  approach as explained in appendix B.
The prediction error between the data and the predicted waves was expressed in 
terms of RMS values. Since the computation of the weight matrices involves random terms,  
the stability of our results was checked by generating multiple instances of each network type. 
The main findings are summarized in Figs.\,5A and 5B. \\\\


\section*{Appendix B: The \BOSZ model for nearshore wave modeling}
\subsection*{B.1. Background}
The Boussinesq Ocean \& Surf Zone model {\em BOSZ} has been developed to facilitate the assessment 
of nearshore wave processes with a reasonable balance between accuracy and computational cost. 
The solution of a depth-integrated Boussinesq-type equation returns the free surface elevation 
and horizontal flow velocities in the two-dimensinal horizontal plane, 
which leads to a significant reduction in computational effort.
The phase-resolving nature of the model allows for the solution of wave-by-wave processes 
and provides a platform to study nonlinear nearshore wave processes on the order of 
a wave scale such as shoaling, refraction, wave breaking, wave run-up, and recirculation.

The idea behind a Boussinesq-type approach is the removal of the vertical velocity component 
of the governing equations through depth-integration while retaining low-order accuracy 
of the vertical variation in pressure. This pseudo-3D solution is often sufficient for nearshore waves 
which are mostly driven by the hydrostatic pressure, i.e. the water depth. 
The solution of a Boussinesq-type equation can be obtained in a fraction of time in comparison 
to a full 3D equation with reasonable accuracy for all major processes on the scale of individual waves. 
 
In the present study, the {\em BOSZ} model was employed for the calculations 
of $123$ hypothetical wave scenarios, which provide the foundation for the training process 
of the stdESN and audESN. The $123$ cases are representative for sea states commonly encountered 
along coastlines worldwide.

\subsection*{B.2. Wave scenarios for training of stdESN and audESN}
The {\em BOSZ} model is used to populate a database of synthetic wave regimes based 
on characteristic values of peak periods, $T_p$, and significant wave heights, $H_s$. 
The peak period is the inverse of the frequency, at which a wave spectrum has its maximum value. 
The significant wave height is defined as the average of the highest $1/3$ 
of all individual waves equal $4\sqrt{m_0}$, with $m_0$ denoting the zeroth moment of a wave spectrum.

Large swell waves with long peak periods are produced by strong winds over a long fetch. 
The wave regime is irregular and chaotic in the generation area, i.e. where a storm occurs. 
Subsequently, the dispersion relation controls the propagation of the individual waves 
in dependence of their wave length and period. This causes a process of sorting as long waves 
travel faster than short waves. Consequently, a sequence of individual waves observed far away 
from their generation area will look very much alike, whereas waves in closer proximity 
to a storm will differ in terms of their period.  Low amplitude and long period waves 
are often a characteristic sign of forerunners as they get to a distant location before 
the main swell package arrives. It is natural that the waves at the coast exhibit different 
peak periods and wave heights  depending on the location of the wind field.

In this study, we use $123$ realistic combinations of $H_s$ and $T_p$ as input into
an empirical Pierson-Moskowitz function. This produces sets of plausible distributions
of power spectral density, which are then converted to a series of wave amplitudes with
different periods as input for the {\em BOSZ} model. The wavemaker source term in {\em BOSZ}
then produces an irregular free surface and velocity evolution, which resemble the spectral
distribution defined by the Pierson-Moskowitz function. 

In general, a wide range of combinations of $H_s$ and $T_p$ is possible. We have selected
a range from $8$ s to $20$ s for the peak periods and from $0.5$ m to $6.5$ m for the significant wave heights.
A total of $123$ individual model runs was then computed over a one-dimensional straight slope of $1/20$.
The computational domain was $4000$ m long and composed of $800$ grid cells with $5$ m cell size.
The water depth ranges from $30$ m offshore to $5$ m.

A series of wave gauges is placed along the slope ranging from  $1000\,$m to $3500\,$m 
in intervals of $50$ cells. The first gauge is at the toe of the slope and last gauge is at the end of slope.
The depth changes by $2.5\,$m between every gauge. However, in the current study, only the
last gauge is used which is located at the near-end of the slope in $5$ m depth.
The first and last $50$ cells of the domain are used for a sponge layer to absorb the outgoing waves
in both directions, i.e. no run-up on dry bed is caluclated to avoid reflections. 
The wavemaker source is centered at $500\,$m.
The computed time for each scenario was $2$ hours and the output at the wave gauges 
was saved at a $1$ Hz sampling rate.

The slope leads to typical nonlinear processes such as wave shoaling and generation of higher harmonics, 
which cause a deformation of the initially generated linear waves in terms of vertical and horizontal asymmetry.
For some of the events wave breaking occurs towards the end of the slope in shallow water.

Most numerical computations depend significantly on input and boundary conditions. In our model, 
the wavemaker builds on the theoretical concept presented in Wei et al. \cite{39}. 
The idea behind this approach is to decompose an amplitude spectrum into multiple individual 
monochromatic waves and then generate every single one periodically with a random phase. 
We start off with the definition of the empirical Pierson-Moskowitz spectrum, 
for which only $H_s$ and $T_p$ have to be known. 
The lowest frequency is $\frac{1}{30}\,$Hz which is commonly used as limiting frequency 
of the gravity wave spectrum. The highest frequency depends essentially on the dispersion 
properties of the numerical model and its governing equation. Many Boussinesq-type equations,
such as the one found by Nwogu \cite{37} which is used here, are accurate for 
non-dimensional wavenumbers up to around $\pi$. 
The solution does not necessarily break down for shorter waves but starts developing errors 
in the waves' celerity. We therefore truncate the spectrum at $kh=\pi$, 
where $k$ is the wavenumber and $h$ is the local (offshore) depth.
This cut-off corresponds to a frequency of $\frac{1}{6.21}\,$Hz, i.e. no shorter wave is initially generated 
but could theoretically occur during the computation from nonlinear interactions. 
The spectrum is then divided into equally-spaced frequency bins between $\frac{1}{30}\,$Hz 
and $\frac{1}{6.21}\,$Hz. 
The width of the frequency depends essentially on the computed time of the model run. 
It is crucial that the time series resulting from the superposed monochromatic waves does 
not repeat over the course of the computation. Otherwise artificial wave groups can contaminate the solution. 
With $2\,$hours of computed time, the frequency interval $\Delta f$ for binning is $0.000139\,$Hz. 
Uniform subdivision of the spectrum between the highest and lowest frequencies results in 
$921$ frequency bins. Logically, a longer or shorter computed time would not affect 
the range of frequencies but the frequency bin width and consequently the total number of waves 
in the generation. The wave phases are initially set as random and used consistently 
across all $123$ model runs. Once the waves are generated, 
they move away from the center of the source in both directions. 
The offshore propagating waves are immediately absorbed by a sponge layer 
that mimics an open ocean boundary condition and also serves to absorb 
potential reflected waves from the slope.
\begin{figure}[h!]
	\centering
		\includegraphics[width=0.72\linewidth, angle=-90]{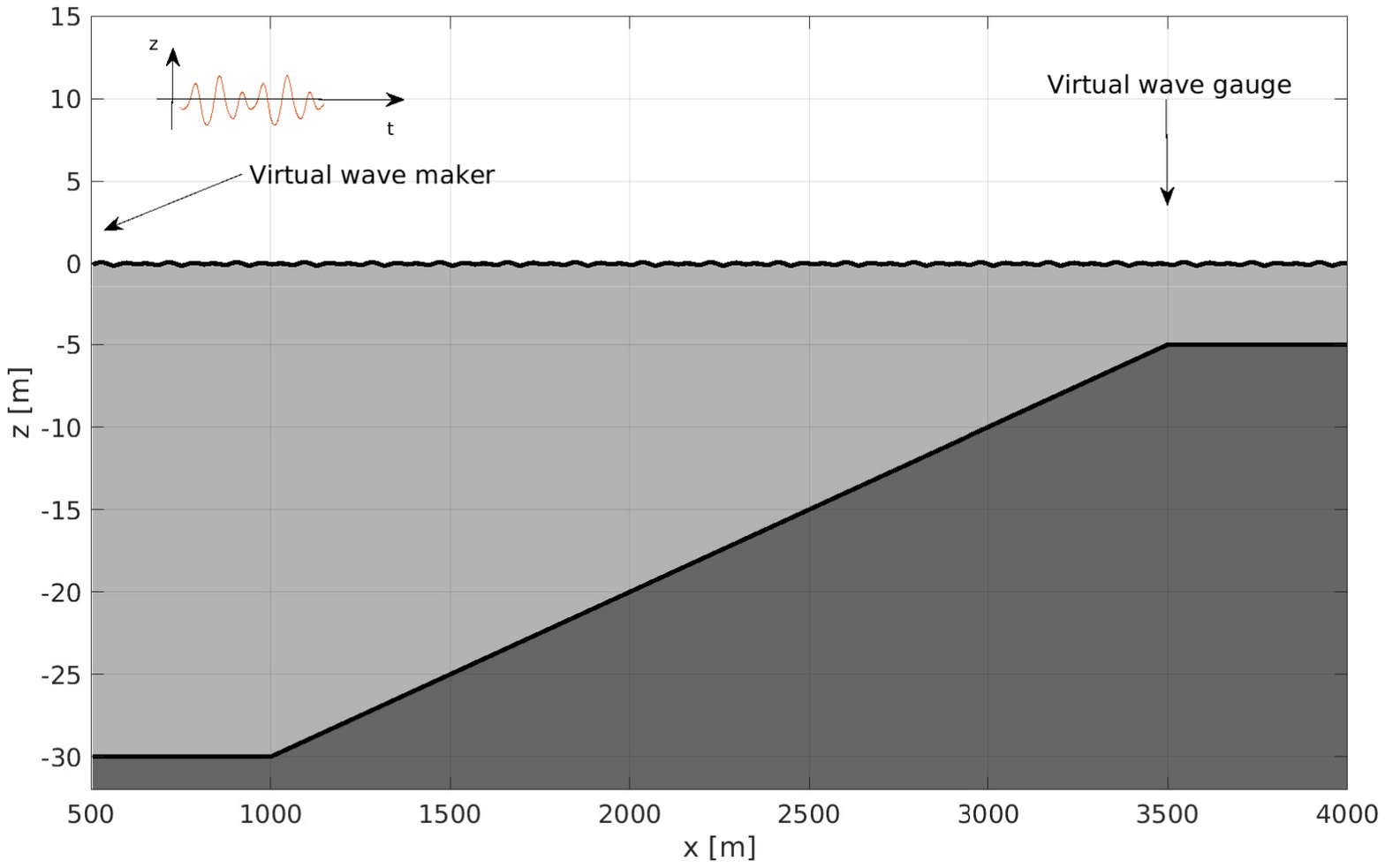}
		\caption{
                          Schematic of the setup for the ocean wave simulations.
                          The numerical domain has a length of $4000\,$m, the wavemaker is
                          placed at a distance of $500\,$m from the left boundary. 
                          The toe of the slope is at $1000\,$m. The wave gauge used in this study
                          was at the top of the slope, at a distance of $3500\,$m from the left boundary.
                    }
		\label{Fig1a}
\end{figure}

\subsection*{B.3. Governing Equations}
The {\em BOSZ} model is built around the long-established Boussinesq-type equations by Nwogu \cite{37}. 
These governing equations contain the Nonlinear Shallow Water (NLSW) equations as a subset 
and include several dispersion terms to account for the non-hydrostatic pressure effects under periodic waves. 
The vertical gradient of the horizontal velocity at a reference depth $z_{\alpha}$ 
is expressed through a truncated Taylor series expansion in combination 
with the vertical irrotationality conditions $u_z = w_x$, $v_z = w_y$. 
These conditions are valid for depth-integrated solutions 
since no overturning of the free surface is possible. This allows for a quadratic approximation 
of the vertical velocity profile expressed in terms of only horizontal velocity components. 
The dispersion properties of the resulting set of equations agrees very well 
with Airy wave theory as long as the wavelength is larger than 
twice the local water depth - a condition mostly given for nearshore waves. 

Like most Boussinesq-type and non-hydrostatic equations the Nwogu equations \cite{37} 
were derived from the Euler equations of motion and consequently lead to a formulation 
of the momentum equations with fluxes and local acceleration variables in non-conservative form. 
This does not necessarily pose an immediate problem for continuous functions, 
i.e. as long as the solution of free surface and velocity is smooth and free of abrupt transitions.

Obviously, Boussinesq-type and non-hydrostatic equations were primarily developed 
to provide solutions for nearshore processes - a regime where wave breaking is common. 
The presence of wave breaking creates a flow field with both subcritical as well as supercritical regimes. 
At the same time, wave breaking leads to sudden transitions in the form of shocks. 
The classical NLSW equations with their hyperbolic structure cater to the solution 
of shocks and flow discontinuities, which can be seen analogous to breaking waves. 
However, a correct shock solution requires that the flux terms are formulated 
in terms of conserved variables. This changes the units in the momentum equation 
terms to $[\mathrm{m}^2 \,/ \, \mathrm{s}^2]$ instead of $[\mathrm{m}\,/\, \mathrm{s}^2]$ 
and supports the mass transport across 
flow discontinuities. Subsequently, Roeber et al. \cite{33} adopted this concept 
and reformulated the equations in conserved variables. 

The equations express a balance between flux and dispersion with additional source terms. 
They can be cast in vector form and are shown here in $x$-direction only as
\begin{equation}
\label{eq:BZ1} 
\text{U}_t + \mathscr{F}(\text{U})_x - \mathscr{S}_{bed} = - \mathscr{S}_{dsp} + \mathscr{S}_{frc} + \mathscr{S}_{wbr} + \mathscr{S}_{wmk} 
\end{equation} 
where $\text{U}$ is the vector of conserved variables, $\mathscr{F}(\text{U})$ is the flux vector, 
$\mathscr{S}(\text{U})_{bed}$ is the bed slope source term, $\mathscr{S}(\text{U})_{dsp}$ 
includes the dispersion terms with spatial derivatives, $\mathscr{S}(\text{U})_{frc}$ 
accounts for bottom friction, $\mathscr{S}(\text{U})_{wbr}$ is a dissipative term 
to account for breaking waves, and $\mathscr{S}(\text{U})_{wmk}$ 
represents local wave generation analogous to a wavemaker paddle in a laboratory environment.


The vectors are given in differential form as
\begin{equation}
\label{eq:BZ2} 
    \text{U} = 
    \left[ \begin{array}{c}
    H \\
    P\end{array} \right]
    \hspace{1cm}
    \mathscr{F}(\text{U})
    =
    \left[ 
    \begin{array}{c}
    Hu \\
    Hu^2+\frac{1}{2}g \eta^2 + g \eta h
    \end{array} 
    \right]
    \hspace{1cm}
    \mathscr{S}(\text{U})_{bed} = 
    \left[ \begin{array}{c}
    0 \\
    g \eta h_x \end{array} \right]
\end{equation}
\begin{equation}
\label{eq:BZ3} 
        \mathscr{S}(\text{U})_{dsp} = 
    \left[
    \begin{array}{c}
    \psi_C \\
    u\psi_C + H_t\psi_P
    \end{array} 
    \right]
    \hspace{4.6cm}
    \mathscr{S}(\text{U})_{frc} = 
    \left[ 
    \begin{array}{c}
    0\\ 
    \frac{g n^2 u^2} {H^{1/3}}
    \end{array} 
    \right]
\end{equation}
\begin{equation}
\label{eq:BZ4} 
	\mathscr{S}(\text{U})_{wbr} = 
    \left[ \begin{array}{c}
    0  \\
    \left(\nu_t H {u}_x\right)_x 
    \end{array} \right]
    \hspace{1cm}
    \mathscr{S}(\text{U})_{wmk} = 
    \left[ \begin{array}{c}
    \sum\limits_{i=1}^{M_\omega}  D_{i}\hspace*{1 mm} \text{cos}\left[k_i\left(x \right) - \omega_i\hspace*{1 mm}  t + \phi_{i} \right]  \\
    0 
    \end{array} \right].
\end{equation}
Here, $H$ and $h$ are the total flow depth and still water depth, respectively. 
$u$ is the horizontal velocity, $g$ is gravity, $\eta$ denotes the free surface, 
$n$ is the Manning roughness coefficient of units $[\mathrm{s}\,\mathrm{m}^{-1/3}]$, 
and $\nu_t$ is the eddy viscosity necessary for wave breaking closure. 
The wavemaker source term generates an oscillating superposition 
of individual monochromatic waves each with a random phase $\phi$, 
magnitude $D$, wave number $k$, and angular frequency $\omega$, over the entire computed time.

The local acceleration term in the momentum equation accounting for frequency dispersion 
with mixed space-time derivatives is given by
\begin{equation}
\label{eq:BZ5} 
P = Hu + H\left[\frac{z_\alpha^2}{2}\left(u_{xx}\right) +z_\alpha(hu)_{xx}\right]
\end{equation} 
and the dispersion terms with only spatial derivatives are denoted by
\begin{equation}
\label{eq:BZ6} 
\psi_C = \left[ \left( \frac{z_\alpha^2}{2} - \frac{h^2}{6} \right) h u_{xx} + \left(z_\alpha + \frac{h}{2}\right) h (hu)_{xx} \right]_x ,
\end{equation} 
\begin{equation}
\label{eq:BZ7} 
\psi_{P} = \left[\frac{z_\alpha^2}{2} u_{xx} + z_\alpha (hu)_{xx} \right] ,
\end{equation}
where $z_\alpha$ is the reference depth, at which the horizontal velocity is evaluated. 
The position influences the disperison properties where values closer to the free surface 
favor the accuracy of short wave disperison and vice versa. Since the free surface 
is the result of a superposition of multiple individual waves, 
the reference depth has to work for a wide range of wavelengths 
and cannot be set simply for only one single frequency. 
A value around mid-depth such as $z_\alpha = -0.531 h$ provides optimized shoaling 
and dispersion properties for a range of waves within $kh<\pi$.

%

\subsection*{B.4. Numerical Solution}
The accuracy of the solution of the present Boussinesq-type formulation, 
with the NLSW equations as subset, benefits from conservative numerical methods. 
Therefore, the governing equations are solved on a collocated grid with a combination 
of a Finite Volume scheme for the hydrostatic part of the equations and 
a central differential Finite Difference scheme for the non-hydrostatic pressure correction terms. 
The Finite Volume scheme solves for the spatial gradients of the fluxes, $\mathscr{F}(\text{U})$, 
and bed slope term, $\mathscr{S}(\text{U})_{bed}$, 
based on a solution from the HLLC approximate Riemann solver at all cell interfaces. 
A total variation diminishing method is used as reconstruction method for the values 
of water depth and flow speed as input into the Riemann solver at all grid cell interfaces. 
As higher-order methods reduce numerical diffusion, a $5$th-order reconstruction method is used.

A two-step Runge-Kutta method is then employed for the time integration, 
in which the evolution variables $\text{U}_t$ in Eq. \eqref{eq:BZ1} are discretized as
\begin{align}
\text{W} &= \text{U}^n + \Delta{t} \mathscr{L}(\text{U}^n) \\
\text{U}^{n+1} &= \frac{1}{2}(\text{U}^n + \text{W}) + \frac{\Delta{t}}{2} \mathscr{L}(\text{W}) ,
\label{eq:BZ8} 
\end{align}
where $\mathscr{L}(\text{U})$ accounts for the variation of $\text{U}$ 
in all $\mathscr{F}(\text{U})$ and $\mathscr{S}(\text{U})$ terms of Eq. \eqref{eq:BZ1} 
at the present time step $n$. The auxiliary variable $W$  denotes the predicted values 
of $\text{U}$ which are subsequenty used in the corrector step as $\mathscr{L}(\text{W})$ 
to obtain the values at $\text{U}^{n+1}$ with second-order accuracy in time. 
The time step $\Delta{t}$ varies dynamically depending on the present flow conditions 
in all grid cells of index $i$ as
\begin{equation}
\label{eq:BZ9} 
\Delta{t} = \frac{\Delta x}{2 \cdot \text{max}(u_i^n+\sqrt{gh_i})}.
\end{equation}
Eq. \eqref{eq:BZ9} implies the use of a Courant number of $0.5$. The evolution variable 
in the momentum equation of Eq.$\,\,\,$\eqref{eq:BZ1} containes combined flux and dispersion terms 
together with mixed space and time derivatives. Once the evolution variables are obtained 
at the next time step ($n+1$), a tridiagonal linear systems of equations 
is solved for the horizontal flow velocity $u$ as the solution vector. 
This concludes the calculation at each time step. 
The procedure is repeated until the final time is reached.


\subsection*{B.5. Wave breaking closure}
\label{eq:TKE}
The source terms in Eq.$\,\,$\eqref{eq:BZ4} of the governing equations 
include a dissipative term for wave breaking $\mathscr{S}(\text{U})_{wbr}$,
without which a converging solution would not be possible. 
It is obvious that a depth-integrated solution
cannot produce overturning of the free surface; 
for that a vertical discretization would be necessary.
The closest solution to overturning of the wave front is a discontinuity. 
However, most Boussinesq-type systems do not per se allow
for the formation of discontinuities since the presence of elliptic dispersion terms 
eliminates the purely hyperbolic structure of the underlying NLSW equations. 

Often, solutions for breaking waves can be obtained by making use of some inherently present numerical diffusion.
However, grid refinement ultimately reduces numerical diffusion over discrete grid cells and the solution tends
to the formation of singularities at the position of the wave front. 
This phenomenon is mathematically and numerically correct; 
however, it is not the desired solution for applications.

A dissipative term added to the momentum equations can provide an efficient way 
to reach closure of the problem. Here, we utilize an approach which makes use 
of the eddy viscosity concept, 
a technique which provides a reasonable basis for local dissipation at the breaking wave front. 

The diffusive term in the momentum equation of $\mathscr{S}(\text{U})_{wbr}$ works 
in a similar fashion as the bottom friction term
- though in a more complex way as it involves the time-varying eddy viscosity $\nu_t$ 
which controls the magnitude of the diffusion term.
The magnitude of the dissipative term not only depends on the local flow variables ($H$ and $u$), 
but also on the entire turbulent flow field represented by $\nu_t$. 
It is therefore necessary that $\nu_t$ has a physical basis, since a fixed coefficient does not account
for the variations in the flow regime. 

Nwogu \cite{38} has developed an approach to determine the quantity of turbulent kinetic energy (TKE) 
from a one-equation model that can be used as proxy or the assessment of $\nu_t$ in each time step. 
This technique was later revisited by Kazolea and Ricchiuto \cite{40}, 
and it is further optimized in the present work.

A relation between TKE and $\nu_t$ was defined by Pope \cite{41} in the form
\begin{equation} 
\nu_t = C_\nu \ell_t \sqrt{k},
\label{eq:TKE1}
\end{equation} 
where $\ell_t$ is a coefficient representing the turbulence length scale 
which is here given by the local water depth, $\ell_t=h$. 
The coefficient $C_\nu$ is set to $0.55$ as suggested by Kazolea and Ricchiuto \cite{40}.
The solution of Eq. \eqref{eq:TKE1} then requires an estimate of TKE, 
for which Pope \cite{41} had formulated a general governing equation as
\begin{equation} 
k_t =  - \mathscr{A} - \mathscr{E} + \mathscr{P} + \mathscr{D},
\label{eq:TKE3}
\end{equation} 
where $\mathscr{A}$, $\mathscr{E}$, $\mathscr{P}$, $\mathscr{D}$ denote advection, elimination/destruction, 
production, and diffusion of TKE, denoted by $k$ with units of $[\mathrm{m}^2 \, / \, \mathrm{s}^2]$.
$\mathscr{E}$ and $\mathscr{D}$ are generally rather small quantities and most of the solution of TKE
is controlled by the production term $\mathscr{P}$.

We assume that turbulence is produced where wave breaking occurs. Since overturning of the free surface
is not possible in depth-integrated equations, finding a robust definition of wave breaking is notoriouly difficult.
Here, we make use of the advantage that the governing equations (Eq.$\,\,$\eqref{eq:BZ1})
allow for a reconstruction of the horizontal velocity
at the free surface as
\begin{equation}
\begin{aligned} 
     u\vert_{\eta} = u + \frac{1}{2}\left(z_\alpha^2-\eta^2\right)u_{xx}+\left(z_\alpha-z\right)\left(hu\right)_{xx}.
\end{aligned}
\label{eq:TKE4}
\end{equation}
It is then assumed that wave breaking begins once the flow at the wave crest becomes supercritical, 
i.e. the horizontal velocity at the wave crest exceeds the wave celerity \cite{42}. 
Therefore, $\mathscr{P}$ is calculated locally where the criterion of $Fr \geq 1$ is met.
The other terms in Eq.$\,\,$\eqref{eq:TKE3} are computed throughout the entire domain 
and lead to a distribution of TKE across the surf zone.
Consequently, the diffusive term $\mathscr{S}_{wbr}$ is largest at the wave breaking front. 
As breaking waves move to shore,
some TKE is left behind the breaking crests and merges with TKE produced by subsequently breaking waves.
This can be seen analogously to white water and foam production in a dissipative surf zone.
The distribution of TKE helps with the convergence of solutions over a wide range of grid cell sizes 
without the occurrence of singularities. 

The terms in Eq.$\,\,$\eqref{eq:TKE3} are detailed by Nwogu \cite{38} and involve several empirical coefficients. 
Here, we express a modified version of Eq.$\,\,$\eqref{eq:TKE3} without the need for case-dependent tuning.
The TKE advection term is conventionally defined as
\begin{equation} 
\mathscr{A} = u k_{x} .
\label{eq:TKE_A}
\end{equation}
The TKE elimination/destruction term with the coefficient $C_d=(C_v)^3$ is denoted as
\begin{equation} 
\begin{aligned} 
       \mathscr{E} = C_d \frac{k^{3/2}}{\ell_t}
\end{aligned}
\label{eq:TKE_E}
\end{equation}
and the TKE diffusion term $\mathscr{D}$ is given by
\begin{equation} 
\begin{aligned} 
       \mathscr{D} = \nu k_{xx},
\end{aligned}
\label{eq:TKE_D}
\end{equation}
where $\nu$ is the kinematic viscosity of water. 

The last term necessary for closure is the term defining production of TKE. 
As mentioned above, it is only computed in cells where the threshold $Fr \ge 1$ is exceeded,
i.e. the local quantity of $u \vert_\eta$ reaches at least $\sqrt{gh}$. 
The production term of TKE can be expressed in analogy to the Boussinesq eddy viscosity concept
which relates the turbulence shearstresses to the velocity gradients:
\begin{equation} 
\mathscr{P} = \frac{\ell_t^2}{\sqrt{C_d}} \left[u_z\vert_{\eta} \cdot u_z\vert_{\eta} \right]^{3/2}.
\label{eq:TKE_P}
\end{equation}
The vertical gradient of the horizontal velocities can be computed from the truncated Taylor series expansion inherent
to the governing equations in combination with the vertical irrotationality condition as
\begin{equation} 
       u_z\vert_{\eta} = -\eta u_{xx} - (hu)_{xx}.
\label{eq:u_z}
\end{equation}
It is then understood that energetic wave breaking, such as encountered in plunging breakers,
leads to an increase in the vertical gradient of the horizontal velocities under the wave crest, $u_z\vert_{\eta}$
and consequently to a suddden increase in $\mathscr{P}$.
Once all components of the turbulence closure model are computed, 
the governing equation of TKE, Eq.$\,\,$\eqref{eq:TKE3} is integrated in time 
with the same Runge-Kutta scheme as the governing equations, Eq. \eqref{eq:BZ1} 
and the quantity of $\nu_t$ is dynamically determined from Eq.$\,\,$\eqref{eq:TKE1} 
at each time step and in each grid cell.
This provides the variables for the time-varying dissipative term for wave breaking, 
$\mathscr{S}(\text{U})_{wbr}$, in Eq.$\,\,$\eqref{eq:BZ4}.



\end{document}